# POLARIMETRIA ÓPTICA E MODELAGEM DE POLARES OBSERVADAS NO OPD/LNA NO PERÍODO DE 2010-2012

# OPTICAL POLARIMETRY AND MODELING OF POLARS OBSERVED IN OPD/LNA IN THE PERIOD 2010-2012

**Karleyne M. G. Silva**[1]
**Cláudia V. Rodrigues**[1]
**Joaquim E. R. Costa**[1]
**Deonísio Cieslinski**[1]
**Leonardo A. Almeida**[1]
**Victor S. Magalhães**[1]

**RESUMO:** *Neste trabalho apresentamos os primeiros resultados do estudo de uma nova amostra de 7 candidatas a polares a partir de dados polarimétricos obtidos no observatório do Pico dos Dias / LNA. Dos 4 objetos analisados até o momento, confirmamos a presença de polarização alta e variável em 3, o que indica a presença de emissão ciclotrônica e sua classificação como polares. Esses dados serão modelados utilizando-se o código CYCLOPS.*
**Palavras-chave:** variáveis cataclísmicas; estrelas binárias; polarimetria.

**ABSTRACT:** *In this work, we present the first results of a study of a new sample of 7 polar candidates from polarimetric data obtained at the Pico dos Dias / LNA observatory. From the four polars analysed so far, we confirm the presence of high and variable polarization in 3. The data will be analysed using the code CYCLOPS.*

[1] Divisão de Astrofísica - Instituto Nacional de Pesquisas Espacias - INPE. E-mail: karleyne@gmail.com.

## 1. INTRODUÇÃO

As polares (Figura 1) são estrelas binárias onde existe a transferência de plasma entre as componentes do sistema devido à grande proximidade entre elas. O material proveniente da estrela doadora, uma anã vermelha, é capturado pelo intenso campo magnético da estrela receptora, uma anã branca. O material acumulado na superfície da anã branca forma uma região de acréscimo que emite radiação polarizada no óptico e radiação em raios X (CROPPER, 1990). Essa radiação é modulada com o período orbital do sistema.

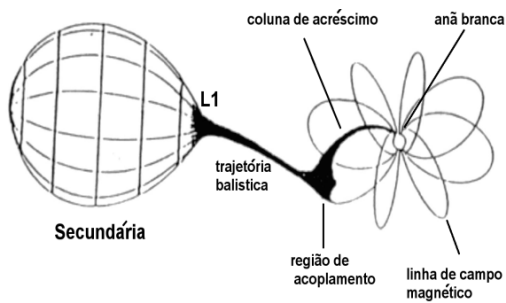

**Figura 1 - Representação de uma polar e descrição das componentes do sistema. Fonte: Adaptada de Cropper (1990).**

Dois catálogos online llistam as polares e candidatas conhecidas: Downes (2001) apresenta cerca de 80 polares confirmadas e Ritter e Kolb (2003) apresenta 135 sistemas, entre candidatos não confirmados e polares. Dessa forma, 55 candidatas precisam de mais informações para sua classificação. A presença de emissão polarizada confirma ocorrência de emissão ciclotrônica da coluna de acréscimo, logo sua detecção confirma a classificação de uma polar. A não detecção de polarização não descarta a classificação como polar, pois há um grupo de sistemas que apresentam valores muito baixos de polarização (AMORIM, 2011).

O estudo de uma polar permite explorar cenários particulares determinados pela configuração geométrica específica de cada fonte, como a existência de uma ou duas regiões emissoras, a ocorrência de auto-eclipses, a ocorrência de absorção por material localizado no plano orbital ou ao longo das linhas do campo magnético e a distribuição de temperaturas e densidades ao longo da região de acréscimo. O estudo de uma amostra homogênea utilizando-se uma mesma metodologia permite determinar os fatores geométricos e ao excluí-los, estudar as propriedades físicas comuns a todas as polares.

## 2. DESENVOLVIMENTO

2.1 Descrição da amostra

Com o objetivo de compreender os processos físicos que ocorrem em polares, nosso grupo da Divisão de Astrofísica-INPE realizou medidas polarimétricas de um conjunto de 7 polares entre 2010 e 2012 realizadas no Observarvatório do Pico dos Dias/ LNA, utilizando o módulo polarimétrico no telescópio de 1.6 m. Estas candidatas a polares são: SWIFT J2319.4+2619, SWIFT J2218.5+1925, V393 Pav, V2301 Oph, CCS100216:1503-2207, 1RXS J1002-192534 e RX J0154.0-5947.

**Tabela 1 - Descrição dos dados de nossa amostra**

| Objeto | Data | Filtro | Tempo de exposição (s) | Número de imagens |
|---|---|---|---|---|
| SWIFT J2319.4+2619 | 10-11/08/2010 | V | 100 | 128 |
|  | 11-12/08/2010 | R | 60 | 144 |
|  | 30/09-01/10/2010 | V | 90 | 83 |
|  | 13-14/10/2010 | V | 90 | 125 |
|  | 20-21/09/2011 | R | 200 | 64 |
| SWIFT J2218.5+1925 | 03-04/11/2010 | V | 60 | 144 |
|  | 14-15/06/2012 | R | 60 | 37 |
|  | 15-16/06/2012 | R | 60 | 96 |
|  | 16-17/06/2012 | I | 60 | 136 |
| V393 Pav | 09-10/08/2010 | I | 80 | 144 |
|  | 10-11/08/2010 | V | 80 | 176 |
|  | 11-12/08/2010 | R | 100 | 128 |
|  | 08-09/10/2010 | R | 90 e 120 | 73 |
|  | 13-14/10/2010 | V | 80 | 96 |
|  | 14-15/10/2010 | I | 60 e 90 | 111 |
| CCS100216:1503-2207 | 15-16/06/2012 | R | 60 | 80 |
|  | 16-17/06/2012 | R | 60 | 100 |
|  | 16-17/06/2012 | V | 60 | 128 |
|  | 17-18/06/2012 | I | 50 | 160 |
|  | 17-18/06/2012 | V | 50 | 129 |
| 1RXS J1002-192534 | 24-25/04/2012 | V | 60 | 176 |
|  | 25-26/04/2012 | I | 90 | 161 |
| RX J0154.0-5947 | 06-08/08/2010 | V | 30 | 97 |
|  | 07-08/08/2010 | V | 30 | 144 |
|  | 08-09/08/2010 | I | 40 | 160 |
|  | 10-11/08/2010 | V | 40 | 53 |
|  | 02-03/11/2010 | R | 30 | 240 |
|  | 03-04/11/2010 | I | 30 | 128 |
| V2301 Oph | 06-07/08/2010 | I | 60 | 96 |
|  | 07-08/08/2010 | I | 60 | 121 |
|  | 08-09/08/2010 | R | 30 e 60 | 160 |
|  | 09-10/08/2010 | I | 50 | 160 |

A Tabela 1 apresenta o resumo das observações obtidas. Esses novos dados polarimétricos permitirão confirmar a classificação como polar. Nos casos confirmados, será realizada a modelagem da região de acréscimo com o código CYCLOPS. O CYCLOPS é um software que permite a modelagem de dados polarimétricos do óptico e de raios X conjuntamente, fornecendo a determinação de parâmetros físicos e geométricos de polares (COSTA; RODRIGUES, 2009; SILVA et al., 2013).

2.2 Polarimetria

O material ionizado capturado pelo campo magnético da anã branca realiza uma trajetória espiral ao longo das linhas de campo magnético. A emissão é polarizada devido ao processo de emissão ciclotrônica. Quando se observar radiação de material em movimento ao longo das linhas de campo magnético, tem-se polarização circular. Quando se observa o movimento perpendicularmente ao campo magnético tem-se polarização linear. À medida que a linha de visada do sistema muda, devido ao movimento próprio da binária, a polarização varia de uma para outra. Como algumas polares apresentam cerca de 50% de emissão polarizada, elas estão entre os objetos mais polarizados do céu.

O polarímetro utilizado nas observações, descrito em Magalhães et al. (1996), possui um prisma de calcita, que divide o feixe incidente em dois feixes com polarização perpendicular entre si e uma lâmina retardadora que altera a fase entre os dois feixes, incluindo uma modulação no sinal recebido quando se modifica o ângulo de posição da lâmina. Conforme apresentado em Rodrigues, Cieslinski e Steiner (1998), a intensidade das duas imagens geradas, ordinária ($I_o$) e extraordinária ($I_e$), relaciona-se com os parâmetros de Stokes I, Q, U e V do feixe incidente através da relação estabelecida por Serkowski (1974):

$$2I_{e,o} = I + Q\cos^2 2\Psi_i \pm U\operatorname{sen} 2\Psi_i \cos 2\Psi_i \pm V\operatorname{sen} 2\Psi_i,$$

onde i varia de 1 a 16 e representa a posição que o retardador assume durante as observações. Os valores $\psi_i$ e $I_{e,o}$ são conhecidos, estes últimos dois medidos a partir de fotometria de abertura das imagens. Os parâmetros Q, U e V são, então, expressos por:

$$Q = \frac{1}{3}\sum_{i=1}^{8} X_i \cos^2 2\Psi_i,$$
$$U = \sum_{i=1}^{8} X_i \operatorname{sen}^2 2\Psi_i \cos 2\Psi_i \text{ e}$$
$$V = \frac{1}{4}\sum_{i=1}^{8}=1\ X_i \operatorname{sen}^2 2\Psi_i.$$

O grau de polarização linear (p) e o grau de polarização circular (v) são dados por:

$$p = \frac{Q^2 + U^2}{I^2} \quad e \quad v = \frac{v}{I}.$$

O ângulo de polarização linear (θ) e a incerteza associada ($\sigma_\theta$) são estimados a partir de:

$$\theta = \frac{1}{2}\operatorname{tg}-1\frac{U}{Q} \quad e \quad \sigma_\theta = 28{,}65\frac{\sigma_p}{p}.$$

### 2.3 Resultados da redução

As Figuras 2 e 3 apresentam as curvas de polarização das 4 polares analisadas até agora. V2301 Oph é uma polar eclipsante. Pode-se ver que sua polarização circular é consistente com zero, considerando as barras de erros (Figura 2, à esquerda). A polarização linear também é baixa, próxima de zero. Os pontos com erros grandes referem-se a medidas feitas durante o eclipse, quando o fluxo do objeto diminui bastante. De nossa amostra, V2301 Oph era o único sistema confirmado como polar via espectroscopia e o estudo em raios X, no entanto, não apresenta polarização. Polares com esse comportamento foram estudadas por Amorim (2011).

V393 Pav apresentou polarização circular alta, variando de -22 a 3% (Figura 2, à direita). Foram cobertos cerca de três períodos orbitais na banda R. A polarização linear é relativamente baixa. A redução de dados de outros filtros permitirá refinar a efemérides desta nova polar.

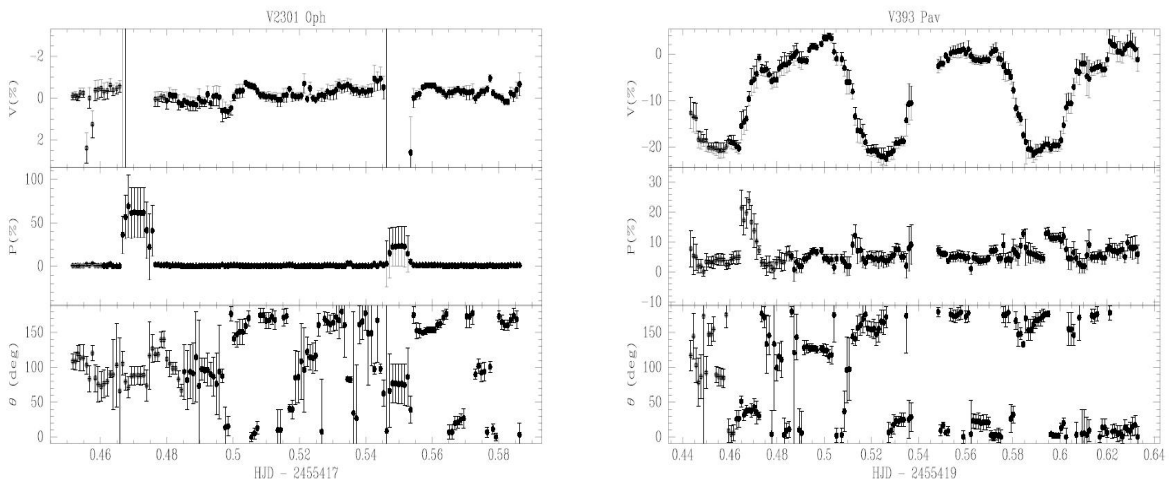

**Figura 2 - Curva de polarização de V2301 Oph e V393 Pav. De cima para baixo, polarização circular (v), polarização linear (p) e ângulo da polarização linear (θ).**

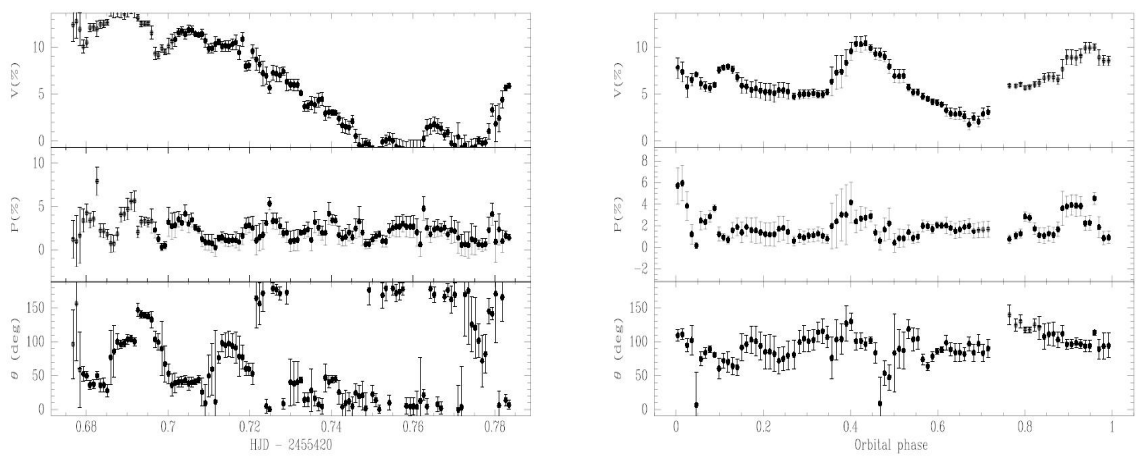

**Figura 3 - Curva de polarização de SWIFT J2319.4+2619 à esquerda e RX J0154.0-5947, à direita. De cima para baixo, polarização circular (v), polarização linear (p) e ângulo da polarização linear (θ).**

SWIFT J2319.4+2619, conforme mostrado na Figura 3 à esquerda, apresenta polarização circular variando de 0 a 15%, os dados que analisamos não cobrem o período orbital completo. A polarização linear é baixa, chegando, no máximo, a 5%.

RX J0154.0-5947, conforme apresentado na Figura 3 à direita, apresenta polarização circular variando de 2 a 13%, nunca chegando a zero. Foi feito um diagrama de fase utilizando-se a efeméride de Ramsay e Cropper (2004). Foi observado, aproximadamente, um período orbital completo. A polarização linear também é baixa, com valor médio de 4%.

## 3. CONCLUSÕES

Apresentamos aqui os primeiros resultados deste novo projeto, que consistiu na polarimetria de uma nova amostra. Verificamos a presença de polarização alta e variável em três das quatro polares analisadas até o momento. Esses resultados inserem-se em um projeto maior para determinação dos parâmetros geométricos e físicos de um grupo de polares de forma homogênea usando o CYCLOPS.

## REFERÊNCIAS